\documentclass[conference, a4page]{templates/IEEEtran}
\pdfoutput=1

\usepackage{graphicx}
\usepackage{subfigure}
\usepackage{url}
\usepackage[nolist,nohyperlinks]{acronym}
\usepackage{framed}
\usepackage{etoolbox}
\usepackage{paralist}
\usepackage{amsmath}
\usepackage{amssymb}
\usepackage{cite}
\usepackage{lipsum}
\usepackage{balance}
\usepackage{tabularx}
\usepackage{tablefootnote}
\usepackage[left=0.62in,top=0.7in,right=0.62in, bottom=0.95in]{geometry}

\graphicspath{{images/}{images/}}

\newcommand{\squeezeup}{\vspace{-3.5mm}}
\newcommand{\squeezeupsmall}{\vspace{-3.0mm}}
\newtoggle{authorcomment}
\toggletrue{authorcomment}

\begin{document}
%

\title{Coexistence of OFDM and FBMC for Underlay D2D Communication in 5G Networks}

\author{\IEEEauthorblockN{Conor Sexton\IEEEauthorrefmark{1},
		Quentin Bodinier\IEEEauthorrefmark{2},
		Arman Farhang\IEEEauthorrefmark{1},
		Nicola Marchetti\IEEEauthorrefmark{1},
		Faouzi Bader\IEEEauthorrefmark{2}, and
		Luiz A. DaSilva\IEEEauthorrefmark{1}} 
	\IEEEauthorblockA{\IEEEauthorrefmark{1}CONNECT - Trinity College Dublin, Ireland,\\}
	\IEEEauthorblockA{\IEEEauthorrefmark{2}SCEE/IETR - CentraleSup\'{e}lec, Rennes, France. \\}
	\IEEEauthorblockA{Email : csexton@tcd.ie\\}}


\maketitle

\begin{abstract}
Device-to-device (D2D) communication is being heralded as an important part of the solution to the capacity problem in future networks, and is expected to be natively supported in 5G.
Given the high network complexity and required signalling overhead associated with achieving synchronization in D2D networks, it is necessary to study asynchronous D2D communications.
In this paper, we consider a scenario whereby asynchronous D2D communication underlays an OFDMA macro-cell in the uplink. 
Motivated by the superior performance of new waveforms with increased spectral localization in the presence of frequency and time misalignments, we compare the system-level performance of a set-up for when D2D pairs use either OFDM or FBMC/OQAM.
We first demonstrate that inter-D2D interference, resulting from misaligned communications, plays a significant role in clustered D2D topologies. 
We then demonstrate that the resource allocation procedure can be simplified when D2D pairs use FBMC/OQAM, since the high spectral localization of FBMC/OQAM results in negligible inter-D2D interference.
Specifically, we identify that FBMC/OQAM is best suited to scenarios consisting of small, densely populated D2D clusters located near the encompassing cell's edge.

\end{abstract}


\section{Introduction}
The need for greater capacity in future cellular networks, and hence more efficient spectrum utilization, has motivated the expected native support of device-to-device (D2D) communication in 5G networks.
Achieving this integration involves a delicate balance between increasing the overall system throughput, and managing the interference imposed by direct transmission between devices. In particular, when utilizing Orthogonal Frequency Division Multiplexing (OFDM) for D2D communications, each D2D pair must be synchronized with the incumbent cellular users (CU) in order to avoid leakage interference between the two. However, the signalling overhead and complexity associated with achieving perfect synchronization by the base station (BS) may be significant, becoming infeasible as larger numbers of D2D pairs are considered. Hence, in this paper, we consider asynchronous D2D communication. That is, we do not assume that D2D pairs are synchronized with either CUs or one another.

There are many works in the literature which aim to mitigate the interference between cellular users and D2D communication (and vice-versa) through resource allocation (RA) and power-loading techniques.
For example, \cite{Feng2013DevicetoDevice} formulates an optimization problem to maximize the sum rate of the D2D users and cellular users, while guaranteeing the quality of service (QoS) requirements for both parties. \cite{Wang2013Resource} aims to maximize the throughput of a D2D link subject to QoS constraints imposed by the cellular users. Most works, however, do not take into account the leakage interference imposed by misaligned D2D communication.
To the best of our knowledge, the inter-D2D interference arising from D2D pairs operating on different resource blocks (RB) is also not well studied in the existing literature. 

Attempting to reduce inter-D2D interference through resource allocation and power-loading techniques is not straight-forward. If inter-D2D interference is considered in the RA and power-loading problems, obtaining the optimal solution in both cases proves to be very difficult, as described in Section III. Furthermore, the viability of such schemes would be questionable, requiring D2D pairs to possess knowledge of the interference contribution from every other D2D pair in the system. Instead, we would like to be able to perform RA and power-loading without needing to account for inter-D2D interference. This would result in schemes that are both simpler and more feasible to implement.

Therefore, we aim to retain the use of less complex resource allocation schemes and reduce inter-D2D interference through other methods.
Given the interest from the community in new waveforms for 5G that possess increased spectral localization, we are motivated to investigate whether inter-D2D interference can be mitigated through the choice of modulation scheme.
In particular, we investigate the use of Filter Bank Multicarrier/Offset QAM (FBMC/OQAM) \cite{Farhang-Boroujeny2011} for D2D pairs. FBMC/OQAM  was chosen due to its high spectral localization and its suitability in scenarios involving asynchronous communications, as well as its prevalence in the literature. Its high spectral containment offers the potential to reduce inter-D2D interference, allowing RA and power-allocation to be accordingly simplified.



This paper considers a scenario in which asynchronous D2D communication underlays an OFDMA macro-cell in the uplink, i.e. D2D pairs reuse the uplink RBs of cellular users.
The D2D users may not be perfectly synchronized in time and frequency, and hence this misalignment introduces interference to/from cellular users and other D2D pairs. We consider a class of D2D applications that result in spatially clustered D2D pairs. 
An example of a scenario belonging to such a class is an automated factory, in the vision of Industry 4.0, operating within the coverage area of a 5G macro-cell.
One consequence of this clustered geometry is that inter-D2D interference may become significant, particularly for small or dense clusters.

We are interested in comparing the system-level performance when D2D pairs use either OFDM or FBMC/OQAM in the scenario outlined in the previous paragraph. Specifically, we are interested in investigating whether FBMC/OQAM offers us an alternative to RA for reducing inter-D2D interference.
We build upon our work in \cite{Bodinier20165G}, which provides interference tables capturing the effects of misaligned D2D users in time and frequency onto OFDMA-based cellular users in the uplink band. We also draw upon the work of \cite{Medjahdi2010a, Bodinier2016ICT} in order to characterize the interference imposed between entities utilizing different waveforms.

The use of FBMC/OQAM in underlay D2D communications has previously been considered in the literature in \cite{xing2014investigation}, where the authors assume that D2D pairs synchronize to CUs only if they are in the same cell. As a result, the most significant interference experienced by D2D pairs is due to the leaked power from either inter-cell CUs or other D2D pairs. In contrast, in this paper, we assume that D2D pairs are not synchronized with the encompassing macro-cell and focus on the intra-cell interference between asynchronous D2D pairs for a clustered topology in a single cell scenario. \cite{Mukherjee2015} also considers the use of a different modulation scheme in D2D communications to reduce inter-D2D interference. However, the authors instead consider Universal Filtered Multi-Carrier (UFMC), and only examine inter-D2D interference between D2D devices in neighbouring cells. They also do not attempt to ascertain which scenarios are best suited to the adoption of the new waveform under their consideration, which is a key focus of this work.


The main contributions of this paper are as follows:
\begin{itemize}
\item We show the effects of inter-D2D interference, taking into account leaked power from sub-bands, for clustered D2D scenarios under various system set-ups.
\item We demonstrate that the optimal RA and power-loading schemes can be simplified when D2D pairs use FBMC/OQAM, as the inter-D2D interference becomes negligible due to the high spectral containment of FBMC/OQAM.  
\item We identify that the adoption of FBMC/OQAM is particularly suited to scenarios consisting of small, densely populated D2D clusters located near the encompassing cell's edge.
\end{itemize}
\section{System Model}
In this paper, we investigate an OFDMA based network in which D2D pairs are permitted to reuse the uplink resources of the incumbent cellular users in an underlay fashion, subject to interference constraints. We stay consistent with the literature and consider uplink resource sharing for two reasons. Firstly, in the uplink, all of the interference imposed by the D2D users onto the cellular users is experienced at the base station (BS), enabling this type of interference to be mitigated through BS coordination. Secondly, and most importantly, some of the pilot information broadcast in the downlink is crucial and should not be interfered with. 

\begin{figure}[t]
  \centering
    \includegraphics[width=2.7in,height=2.7in,clip,keepaspectratio]{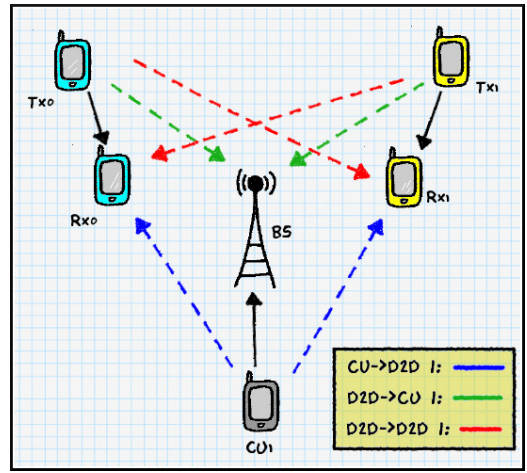}
    \vspace{-1.0mm}
    \caption{Simplified diagram showing two D2D users and one cellular user with both interference channels (dashed lines), and useful channels (solid black lines) outlined.}
    \label{fig:simple_model}
\end{figure}

Fig.~\ref{fig:simple_model} illustrates a simplified scenario in which D2D communication underlays an OFDMA network in the uplink. In a more general scenario, $M$ D2D pairs coexist with $N$ cellular users (CU), and reuse the uplink spectral resources. $C\ =\ \{1,...,N\}$ denotes the set of incumbent (cellular) users and $D\ =\ \{1,...,M\}$ denotes the set of D2D pairs. The useful and interference channels in Fig.~\ref{fig:simple_model}, shown as solid and dashed lines, respectively, are presented in Table \ref{tab:fig1channels}. CUs do not interfere with each other as we assume they are perfectly synchronized by the BS. Therefore, there are three main interference types requiring consideration:
\begin{enumerate}
\item The D2D pairs interfere with the incumbents' transmissions. Since we are investigating uplink resource sharing, this interference is observed at the base station. 
\item Conversely, the incumbents interfere with the D2D pairs at D2D receivers. 
\item D2D pairs interfere with each other (inter-D2D interference).
\end{enumerate}

\bgroup
\def\arraystretch{1.3}
\begin{table}[t]
\caption{Useful and interference channels for Fig.~\ref{fig:simple_model}}
\begin{center}
\begin{tabular}{|c|c|c|}
\hline
\hline
\multicolumn{3}{|c|}{\textbf{Useful Channels}} \\ \hline
$\rm cu_1 \rightarrow eNb$ & $ \rm Tx_0 \rightarrow Rx_0$ & $ \rm Tx_1 \rightarrow Rx_1$  \\ \hline
\multicolumn{3}{|c|}{\textbf{Interference Channels}} \\ \hline
$ \rm cu_1 \rightarrow Rx_0$ & $ \rm Tx_0 \rightarrow eNb$ & $ \rm Tx_1 \rightarrow eNb$ \\ \hline
$ \rm cu_1 \rightarrow Rx_1$ & $ \rm Tx_0 \rightarrow Rx_1$ & $ \rm Tx_1 \rightarrow Rx_0$  \\ \hline
\hline
\end{tabular}
\end{center}
\label{tab:fig1channels}
\vspace{-5.0mm}
\end{table}
\egroup

In each type, we consider both co-channel and adjacent-channel interference. We model D2D pairs using either FBMC/OQAM or OFDM, while the cellular users are restricted to using OFDM.
The reason to consider the use of FBMC/OQAM for D2D operation lies in its high spectral containment and its low sensitivity to asynchronism in the multi-user context \cite{Farhang-Boroujeny2011}, which is expected to decrease the interference both between different D2D pairs and between D2D and cellular users. In this study, we consider FBMC/OQAM transmission based on the PHYDYAS filter \cite{Bellanger}. For more detailed information on the FBMC/OQAM modulation, we refer the reader to \cite{Farhang-Boroujeny2011}.

We consider an OFDMA macro-cell with parameters selected based on the 3GPP LTE standard, as outlined in Table \ref{tab:sim_params} in Section IV. We assume that the cell is fully loaded with each CU assigned a single uplink RB. D2D devices underlay the OFDMA cell by reusing a single uplink RB. In our channel model, we are primarily concerned with pathloss, since we wish to evaluate the performance of both waveforms as we vary several distance related parameters such as cluster size, or distance from the BS. Owing to their popularity in the existing literature \cite{Doppler2009Device, xing2014investigation, Yu2011Resource}, we employ the WINNER II channel models \cite{kyosti2007winner} to provide us with a distance based pathloss, which also incorporates the probability of line-of-sight. Specifically, we use scenario B1 - \textit{urban micro-cell}.


 

\section{Interference Model and Problem Formulation}
\subsection{Interference Model}

The main measure that we base our analysis upon is the SINR experienced by incumbent CUs and D2D pairs. To rate the latter with accuracy, it is necessary to use models of interference that properly estimate the leakage that two asynchronous users inject onto each other. As mentioned earlier, to the best of our knowledge, most studies on D2D underlay operation do not consider leakage between adjacent frequency resource blocks. In papers that do consider leakage, as in \cite{xing2014investigation}, they rely on the Power Spectral Density (PSD)-based model, the shortcomings of which have been demonstrated in \cite{Bodinier2016ICT}.

\begin{figure}[t]
\centering
	\includegraphics[width=\linewidth]{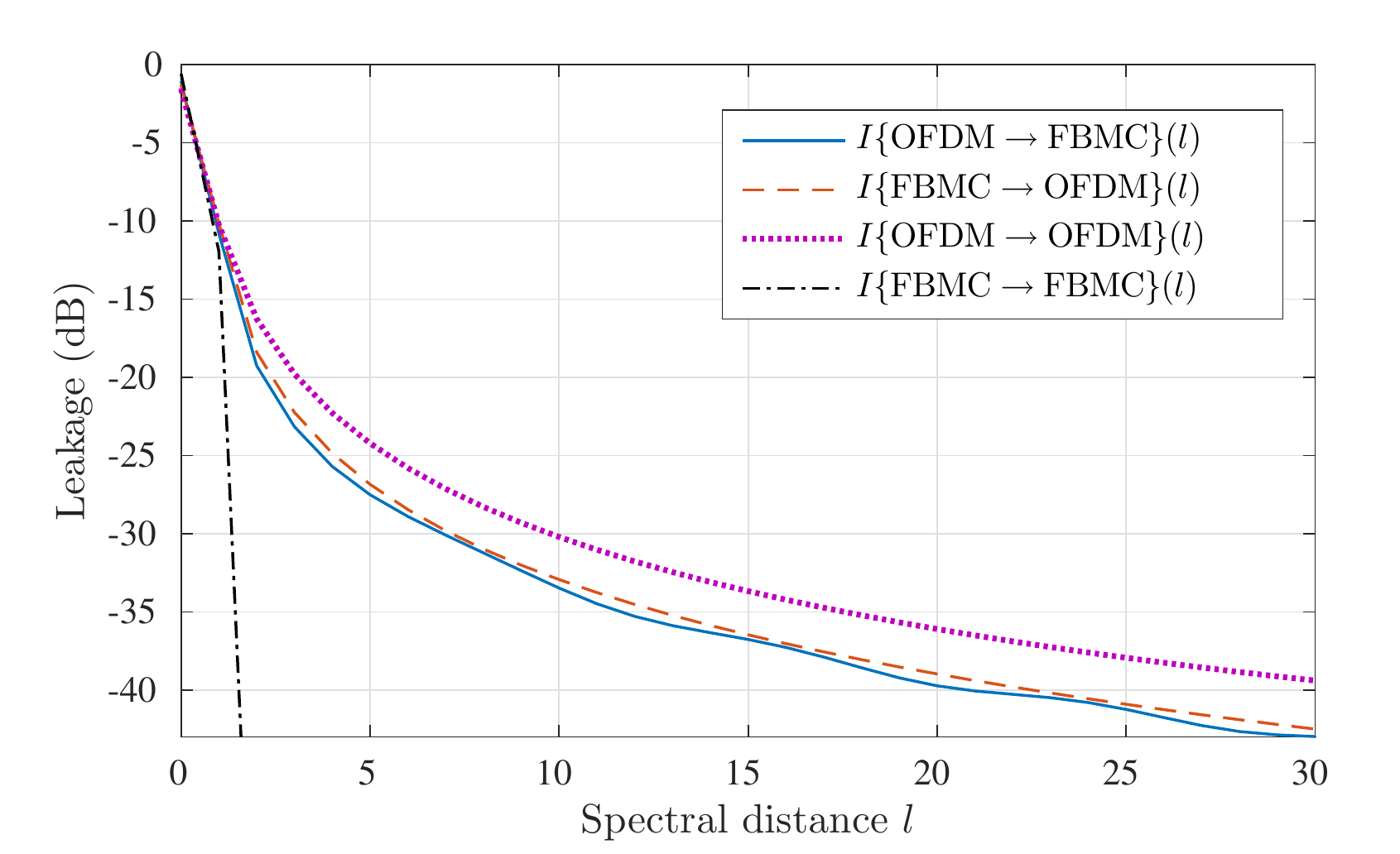}
	\vspace{-6.0mm}
	\caption{Interference tables measuring the value of interference injected between different couples of waveforms according to \cite{Medjahdi2010a, Bodinier20165G, Bodinier2016ICT}.}
	\label{fig:interf_tables}
	\vspace{-5.0mm}
\end{figure}

Fortunately, a number of papers have extensively analysed and precisely modelled the leakage between asynchronous users operating on different parts of the spectrum band, and derived interference tables that we will build our analysis upon \cite{Medjahdi2010a, Bodinier20165G, Bodinier2016ICT}. More precisely, we draw upon the work of \cite{Medjahdi2010a} to rate the interference from FBMC/OQAM to FBMC/OQAM users, or OFDM to OFDM users. Additionally, we obtain the interference from OFDM to FBMC/OQAM and from FBMC/OQAM to OFDM according to the recent works of \cite{Bodinier20165G} and \cite{Bodinier2016ICT}. These works allow us to rate the value of $I\{\text{A}\rightarrow\text{B}\}(l)$, which corresponds to the interference injected by a subcarrier of waveform A to a subcarrier of waveform B at spectral distance $l$.

In this paper, we use the interference table plotted in Fig.~\ref{fig:interf_tables}.
This figure shows that the use of FBMC/OQAM for D2D operation will only marginally reduce the interference between cellular and D2D users, as $I\{\text{OFDM}\rightarrow\text{FBMC/OQAM}\}$ is only slightly less than $I\{\text{OFDM}\rightarrow\text{OFDM}\}$. This has been thoroughly explained in \cite{Bodinier2016ICT}. However, the interference between asynchronous D2D users will be drastically reduced if they use FBMC/OQAM instead of OFDM, since $I\{\text{FBMC/OQAM}\rightarrow\text{FBMC/OQAM}\}$ is considerably lower than $I\{\text{OFDM}\rightarrow\text{OFDM}\}$ . 

\subsection{Optimization Problem Formulation}
We wish to improve the performance of the cellular network using underlay D2D communication. A D2D pair is allowed to transmit when the interference introduced on the incumbent network does not prevent the incumbent CUs from satisfying their minimum SINR constraints. The D2D transmissions affect the SINR experienced at the BS and hence the CUs suffer from adjacent channel interference. The SINR of the CU indexed by $i$ can therefore be expressed as
\begin{equation}
\gamma_i\ =\ \frac{P_ih_{iB}}{\sigma_\nu^2 + \sum_{j \in D} \sum_{r \in \rm R} \sum_{m \in b_r} \omega_{jr} h_{jB} \Omega_{mi}^{\rm D}},
\end{equation}
where $P_i$ is the transmit power of the $i^{\rm th}$ CU, $h_{iB}$ is the channel gain between the $i^{\rm th}$ CU and the BS, $\sigma_\nu^2$ is additive white Gaussian noise variance, $r$ indexes the resource blocks (RB) across the entire band, $m$ indexes the subcarriers in a particular RB band $b_r$, $\omega_{jr}$ is a resource reuse indicator where $\omega_{jr} = 1$ when D2D pair $j$ reuse RB $r$, and $\omega_{jr} = 0$ otherwise. Finally, $\Omega_{mi}^{\rm D}$ is the interference introduced by the $m^{\rm th}$ subcarrier onto the RB used by the $i^{\rm th}$ CU. $\Omega_{mi}^D$ is given by
\begin{equation} \label{omega_1}
\Omega_{mi}^{\rm D}\ =\ \sum_{k \in b_i} \frac{P_{m}}{P_0} I(|k-m|),
\end{equation}
where $k$ indexes the subcarriers in the incumbent band $b_i$ used by user $i$, and $I(|k-m|)$ is the appropriate interference table $I\{\text{A}\rightarrow\text{B}\}$ in Fig.2, depending on the waveform being used by the D2D pairs. 

A D2D receiver will experience two types of interference: i) interference from CUs, and ii) interference from other D2D pairs. The SINR experienced on subcarrier $m$ at the D2D receiver of pair $j$ is given by

\begin{equation} \label{d2dSINR}
\gamma_{jm}\ =\ \frac{P_{jm} h_j}{\sigma_\nu^2 + I_{\rm cu} + I_{\rm D2D}},
\end{equation}
where $P_{jm}$ is the power of the $j^{\rm th}$ D2D pair on subcarrier $m$. $I_{\rm cu}$ is the interference injected on the $m^{\rm th}$ subcarrier from cellular users using OFDM in the incumbent band, and $I_{\rm D2D}$ is the interference from other D2D users.
$I_{\rm cu}$ is defined as

\begin{equation}
I_{\rm cu} = h_{ij} \sum_{i \in C} \Omega_{im}^{\rm C},
\end{equation}
where $\Omega_{im}^{\rm C}$ is the interference introduced by the $i^{\rm th}$ CU onto the $m^{\rm th}$ subcarrier of D2D pair $j$, and is specified in a similar fashion to equation (\ref{omega_1}).


Finally, $I_{\rm D2D}$ is the interference from other D2D links given by
\begin{equation}\label{eq:id2d}
I_{\rm D2D} = \sum_{d \in D, d \neq j} \sum_{r \in \rm R} \sum_{n \in b_r} \omega_{dr} h_{jd} \Omega_{nm}^{\rm D}\ ,
\end{equation}
where $\Omega_{nm}^{\rm D}$ is the interference injected by the $n^{\rm th}$ subcarrier of the $d^{\rm th}$ D2D user onto the $m^{\rm th}$ subcarrier of $j^{\rm th}$ D2D user.

We can now formulate an optimization problem, using the above SINR expressions, in which the objective is to maximize the sum rate of D2D pairs, subject to a minimum SINR constraint for each CU.

\begin{subequations}\label{first:main}
\begin{align}
{\rm P1:} &\max_{P_m, \omega_{jr}} \sum_{j \in D} \sum_{r \in R} \sum_{m \in b_r} \omega_{jr}  \log(1 + \gamma_{jm}) \tag{\ref{first:main}},
\intertext{subject to}
&\omega_{jr} \in \{0,1\}, \forall j,r \label{p1:a},\\
&\sum_{j \in D} \omega_{jr} \leq 1, \forall r \in R \label{p1:b},\\
&\sum_{r \in R} \omega_{jr} = 1, \forall j \in D \label{p1:c},\\
&\gamma_i \geq \textrm{SINR}_{\rm min}^{\rm C} , \forall i \in C \label{p1:d},\\
& P_j = \sum_{m \in b_j} P_m < P_{\rm max}^{\rm D},
\end{align}
\end{subequations}
where $\textrm{SINR}_{\rm min}^{\rm C}$ is the minimum acceptable SINR that a CU must achieve.

Optimization problem P1 is a mixed integer non-linear programming (MINLP) problem from which it is difficult to obtain the solution directly. Accordingly, we split the optimization problem into two sub-problems. First, we perform RB assignment, which is a discrete optimization problem. Once RBs have been assigned, we perform power-loading for the D2D pairs. 

Even after splitting P1 into two simpler problems, solving them remains complicated due to the inclusion of inter-D2D interference. The main source of this complexity lies in the fact that $I_{\rm D2D}$ (\ref{eq:id2d}) is a function of the power assigned to each subcarrier of each D2D transmitter. Therefore, the different variables over which the optimization is performed are coupled, as the SINR of each D2D pair affects the SINR of every other pair, complicating (\ref{first:main}). Furthermore, incorporating inter-D2D interference into the RA scheme would assume that every D2D pair is able to obtain information regarding the interference contribution from every other D2D pair. This is an unrealistic assumption, requiring an exchange of information between D2D pairs before any resource is assigned. Hence, in reality, we would like to be able to perform both RB assignment and power-loading without needing to consider inter-D2D interference.

Therefore, we consider a simplification of P1 (\ref{first:main}) where the SINR $\gamma_{jm}$ (\ref{d2dSINR}) is reduced to
\begin{equation}\label{gamma_prime}
\gamma_{jm}^\prime\ =\ \frac{P_m h_j}{\sigma_\nu^2 + I_{\rm cu}}.
\end{equation}

The effects of inter-D2D interference are not taken into account in (\ref{gamma_prime}). Instead, we are motivated to develop alternative methods to mitigate inter-D2D interference other than through RA, namely through the use of FBMC/OQAM for D2D pairs.
Accordingly, we demonstrate that if D2D pairs use FBMC/OQAM, then there is no significant performance loss incurred by performing RA and power loading without taking into account the inter-D2D interference. This greatly reduces the complexity of the resource allocation (RA) schemes and ensures that the power-loading objective function is convex. It is also more realistic as it makes no assumptions regarding the information a D2D pair possesses about every other D2D pair in the cluster.


Given the above simplifications, the two intermediate problems to be solved can be rewritten as follows.

\subsubsection{RB Assignment}
We assume each cellular user is assigned a single RB and that there are as many CUs as RBs. Since we only consider pathloss in our channel model, RBs can be randomly assigned to CUs. We then want to assign one RB to each D2D pair such that the interference experienced by each D2D pair from the CUs is minimized. The interference experienced by D2D pair $j$ on RB $r$ is given by
\begin{equation} \label{I_1}
I_{jr}\ =\ \sum_{i \in C} \sum_{m \in b_i} \sum_{k \in b_r} \frac{P_{m}}{P_0}h_{ji} I(|k-m|).
\end{equation}
The assignment problem can be specified as follows
\begin{subequations}\label{first:main2}
\begin{align}
 {\rm P2:} &\min_{\omega_{jr}} \sum_{j \in D}  \omega_{jr} \phi_{jr} \tag{\ref{first:main2}}, \\
 \intertext{subject to}
 &\omega_{jr} \in \{0,1\}, \forall j,r,\\
 &\sum_{j \in D} \omega_{jr} \leq 1, \forall r \in R,\\
 &\sum_{r \in R} \omega_{jr} = 1, \forall j \in D,
\end{align}
\end{subequations}
where $\phi_{jr}$ is the interference from CUs experienced by D2D pair $j$ on RB $r$. 
Problem P2 is a combinatorial optimization problem, made complicated by the fact that multiple D2D users may have the same optimal RB assignment. In line with the literature \cite{Feng2013DevicetoDevice,Han2014Bipartite,Wang2014Fast}, we utilize the well-known Kuhn-Munkres algorithm (commonly known as the Hungarian method), to solve the uplink resource assignment problem for D2D pairs.

\subsubsection{Power-loading}
Having assigned an RB to each D2D pair, power-loading can now be performed. The power-loading optimization problem is similar to P1, with the discrete constraints (\ref{p1:a}-\ref{p1:c}), which relate to RB assignment, removed. Since RB assignment has already been performed, the objective function of optimization P1, i.e., equation (\ref{first:main}), can be simplified as follows
\begin{equation} \label{p2_2}
\max_{P_m} \sum_{j \in D} \sum_{m \in b_j}  \log(1 + \gamma_{jm}^\prime).
\end{equation}

The resulting problem is clearly convex and similar to others in the literature, for example \cite{Shaat2010TwoStep}. The solution can be readily obtained using an appropriate software package.
\section{Results}
\begin{figure}[t]
  \centering
    \includegraphics[width=2.8in,height=2.8in,clip,keepaspectratio]{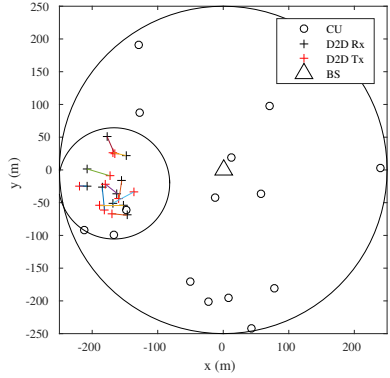}
    \squeezeupsmall
    \caption{Example of clustered scenario consisting of 10 D2D pairs.}
    \label{fig:example_scenario}
\end{figure}

We perform system level simulations to investigate the co-existence of FBMC/OQAM and OFDM. Cellular users are uniformly distributed over the coverage area of the encompassing OFDMA cell. In the clustered scenario, the cluster centre is chosen according to a uniform distribution within the macro-cell area and D2D pairs are uniformly distributed within the cluster area. Fig.~\ref{fig:example_scenario} illustrates an example of a clustered scenario with 10 D2D pairs.

Table \ref{tab:sim_params} lists the key simulation parameters. After distributing both the CUs and D2D pairs within the cell, we then perform RB assignment and power-loading as described in Section III. The average rate per D2D pair is used as the main output metric from simulations. This metric is calculated for two different cases using the SINR expressions described in Section III:
\begin{enumerate}
\item Case 1: D2D pairs use OFDM, CUs use OFDM.
\item Case 2: D2D pairs use FBMC/OQAM, CUs use OFDM.
\end{enumerate}
In both cases, we compare the predicted average rate per D2D pair calculated using  $\gamma_{jm}^\prime$ (which does not take into account inter-D2D interference), with the actual average rate per D2D pair calculated using  $\gamma_{jm}$ (which takes into account inter-D2D interference). 

\bgroup
\def\arraystretch{1.3}
\begin{table}[t]
\caption{Simulation parameters}
\begin{center}
\begin{tabular}{|c|c|}
\hline
\hline
\textbf{Parameter} & \textbf{Value} \\ \hline \hline
inter-site distance (ISD) & 500 m \\ \hline
macro-cell radius & 250 m \\ \hline
carrier frequency & 700 MHz \\ \hline
subcarrier spacing & 15 kHz \\ \hline
number of RBs & {15, 25} \\ \hline
number of CUs & {15, 25} \\ \hline
scenario type & clustered or non-clustered \\ \hline
maximum cluster radius & ISD/5 m \\ \hline
minimum cluster radius & ISD/10 m \\ \hline
maximum D2D Tx. Rx. distance & (cluster radius) $\times$ 2/3 \\ \hline
pathloss model & WINNER II scenario B1 \\ \hline
CU minimum SINR & 10 dB \\ \hline
noise power per subcarrier \tablefootnote{Noise power per subcarrier is calculated using the expression ${-174}\textrm{dBm/Hz}$$ + 10\log_{10}(15\textrm{kHz})$, where ${-174}\textrm{dBm/Hz}$ is the background noise and $15\textrm{kHz}$ is the LTE subcarrier spacing.} ($\sigma_\nu^2$) & -127 dBm \\ \hline
maximum transmit power & 24 dBm \\ \hline
number of iterations & 40000 \\ \hline \hline
\end{tabular}
\end{center}
\label{tab:sim_params}
\squeezeup
\end{table}
\egroup


\vspace{-1.5mm}
\subsection{Effects of inter-D2D interference for both FBMC/OQAM and OFDM}
\begin{figure}[t]
  \centering
    \includegraphics[width=3.3in,height=3.0in,clip,keepaspectratio]{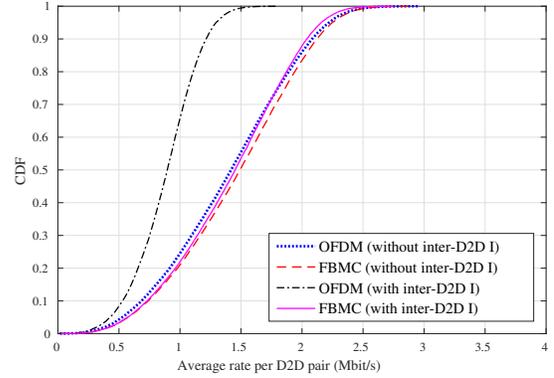}
    \squeezeupsmall
    \caption{Clustered scenario consisting of 10 D2D pairs.}
    \label{fig:basic_2}
    \vspace{-2mm}
\end{figure}

\begin{figure}[t]
  \centering
    \includegraphics[width=3.3in,height=3.0in,clip,keepaspectratio]{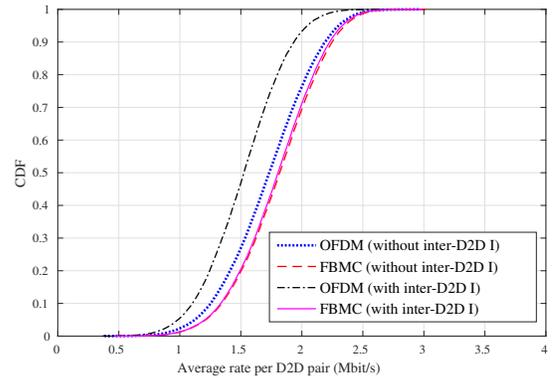}
    \squeezeupsmall
    \caption{Non-clustered scenario consisting of 10 D2D pairs.}
    \label{fig:basic_4}
    \vspace{-5mm}
\end{figure}

\vspace{-1.0mm}
In the first set of results, we show the adverse effects of inter-D2D interference when D2D pairs use OFDM, and how this type of interference may be considered negligible when FBMC/OOAM is instead used. We generate CDFs for the average rate per D2D pair for both the clustered (Fig.~\ref{fig:basic_2}) and non-clustered (Fig.~\ref{fig:basic_4}) scenarios in order to demonstrate the effects of inter-D2D interference for both waveforms.

In the clustered scenario in Fig.~\ref{fig:basic_2}, we observe that when OFDM is used, the gap between the actual and predicted values of rate (calculated using $\gamma_{jm}$ and $\gamma_{jm}'$ respectively) is significant. Conversely, when FBMC/OQAM is used, the actual values of achieved average rate per D2D pair are very close to those calculated without taking inter-D2D interference into account. Thus, FBMC/OQAM provides significant improvement over OFDM by virtue of its inherent ability to mitigate inter-D2D interference. In the non-clustered scenario in Fig.~\ref{fig:basic_4}, we observe that the advantage of using FBMC/OQAM, even when inter-D2D interference is taken into account, is less than the corresponding clustered scenario. This is intuitive, as D2D pairs are farther apart in the non-clustered scenario and, hence, inter-D2D interference does not play such a significant role.

Consequently, we make two observations. First of all, inter-D2D interference plays a significant role in clustered D2D underlay communication. 
Second, we observe that while inter-D2D interference is detrimental to performance when OFDM is used, its effect is negligible when FBMC/OQAM is employed. Therefore, Fig.~\ref{fig:basic_2} and Fig.~\ref{fig:basic_4} show that permitting D2D users to use FBMC/OQAM can significantly facilitate the resource allocation process in the considered scenarios.

\begin{figure}[t]
  \centering
    \includegraphics[width=3.3in,height=3.0in,clip,keepaspectratio]{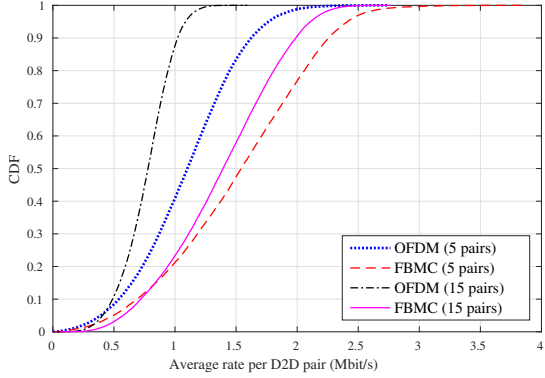}
    \vspace{-2mm}
    \caption{Average rate per D2D pair for different numbers of D2D pairs.}
    \label{fig:basic_5}
\end{figure}

Fig.~\ref{fig:basic_5} reinforces our observations. We display the CDF of the actual average rate per D2D pair, calculated using $\gamma_{jm}$ when inter-D2D interference is considered, in a clustered scenario for 5 D2D pairs and 15 D2D pairs. We first observe that the average rate for both OFDM and FBMC/OQAM decreases as the number of D2D pairs increases, since each D2D transmitter must now use a lower transmit power in order to satisfy the interference constraint specified by (\ref{p1:d}). We also observe that the benefit attributed to using FBMC/OQAM increases as the number of D2D pairs increases, i.e. the gap between the FBMC/OQAM and OFDM curves grows larger as the number of D2D pairs increases. This is due to inter-D2D interference becoming more significant as the number of D2D pairs is increased, despite the fact that pairs must now use less power.

\vspace{-1.5mm}
\subsection{FBMC/OQAM and OFDM performance in varying scenario set-ups}
\vspace{-1.0mm}
In this subsection, we identify that the adoption of FBMC/OQAM is best suited to scenarios consisting of small, dense clusters located far from the encompassing cell's BS. This analysis was performed by varying three key scenario parameters (number of D2D pairs per cluster, cluster radius, and cluster distance to BS), and comparing the relative performance of both waveforms.

\begin{figure}[t]
  \centering
    \includegraphics[width=3.3in,height=3.0in,clip,keepaspectratio]{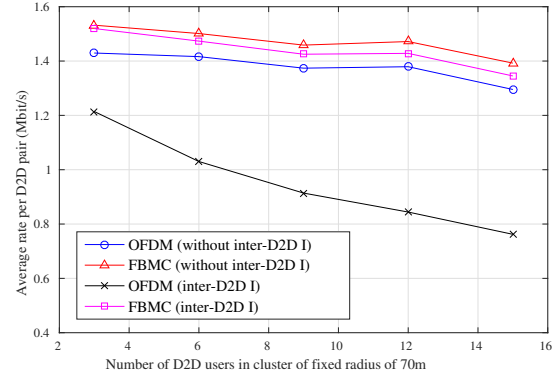}
    \vspace{-2mm}
    \caption{Average rate per D2D pair versus number of D2D users in a cluster of fixed radius of 70m.}
    \label{fig:cluster_density}
    \vspace{-2mm}
\end{figure}

\begin{figure}[t]
  \centering
    \includegraphics[width=3.3in,height=3.0in,clip,keepaspectratio]{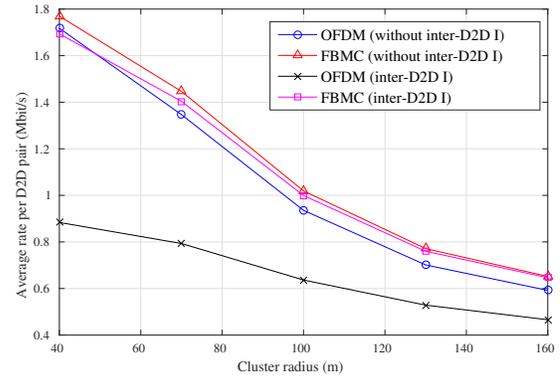}
    \vspace{-2mm}
    \caption{Average rate per D2D pair versus cluster radius for a fixed number of D2D pairs.}
    \label{fig:cluster_radius}
    \vspace{-4mm}
\end{figure}

We first examine the effect that cluster density has on the relative performance of both waveforms. Cluster density can be varied in two manners: i) the cluster radius can be held constant and the number of D2D pairs in the cluster can be altered, or ii) the number of D2D pairs can be fixed and the cluster radius can be altered. We point out that a small cluster with few D2D pairs and a large cluster with many D2D pairs may have similar densities but represent two different types of scenario. Hence, we investigate the effects of density for both of the aforementioned cases (Figs.~\ref{fig:cluster_density} and \ref{fig:cluster_radius}). 


Fig.~\ref{fig:cluster_density} shows the effect of varying the density of a cluster by fixing the cluster radius at 70 m and varying the number of D2D pairs in the cluster. As the cluster density is increased by adding additional D2D pairs, the throughput gain that FBMC/OQAM offers over OFDM also increases. This is intuitive, as inter-D2D interference can be expected to play a more significant role in dense clusters. FBMC/OQAM is, therefore, most effective in dense clusters, consisting of many users, in which inter-D2D interference has a significant impact on the SINR of D2D devices.



Fig.~\ref{fig:cluster_radius} demonstrates the effect of cluster size on the average rate per D2D pair. As the cluster radius is decreased, we observe that the D2D pairs are able to achieve higher rates as the pathloss between a D2D transmitter and receiver also decreases. We also note that the inter-D2D interference becomes more significant, evident by the increasing gap between the curve representing the rate achieved using OFDM when inter-D2D is taken into account, and the rate achieved using FBMC/OQAM. At a cluster radius of 40 meters, the benefit obtained from using FBMC/OQAM is significant. In effect, we are varying the cluster density by fixing the number of D2D pairs, while changing the cluster radius. 
Based on the results presented in Figs.~\ref{fig:cluster_density} and \ref{fig:cluster_radius}, we conclude that FBMC/OQAM is best suited to scenarios consisting of small, dense clusters.

\begin{figure}[t]
  \centering
    \includegraphics[width=3.3in,height=3.0in,clip,keepaspectratio]{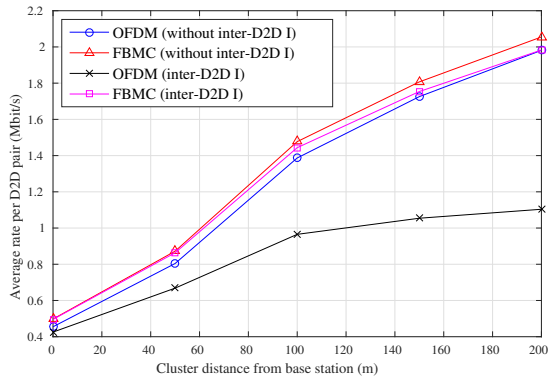}
    \vspace{-2mm}
    \caption{Average rate per D2D pair versus distance from cluster centre to BS.}
    \label{fig:cluster_distance}
    \vspace{-4mm}
\end{figure}

We make two observations with respect to Fig.~\ref{fig:cluster_distance}, which demonstrates the effect of varying the distance from the cluster centre to the BS. First, we note that the average D2D rate increases as the cluster distance from the BS increases. Indeed, since the interference imposed by D2D pairs onto CUs is observed at the BS, D2D transmitters belonging to clusters that are at a greater distance from the BS are permitted to use higher transmit powers. Second, we note that the benefit of using FBMC/OQAM is greater when the cluster is near the macro-cell edge. This can be attributed to the fact that inter-D2D interference increases, and hence becomes more significant, when D2D pairs are permitted to transmit using higher power values. 
It should be reiterated that these results were obtained for a single cell scenario. In a multi-cell scenario, we would expect clusters located near the macro-cell edge to experience increased interference from neighbouring cells.
\section{Conclusion}
In this paper, we considered a scenario whereby asynchronous D2D communication underlays an OFDMA macro-cell in the uplink. We first demonstrated that inter-D2D interference is significant for applications of D2D that result in clustered geometries. Given the suitability of new waveforms for scenarios involving asynchronous communications, we investigated a scenario in which D2D devices use either FBMC/OQAM or OFDM, in coexistence with the encompassing OFDMA macro-cell. 
We demonstrated that the use of FBMC/OQAM alleviates the need to develop more complex RA schemes, owing to its high spectral localization and resulting ability to mitigate inter-D2D interference. 
More precisely, we showed that if D2D pairs use FBMC/OQAM, then there is no significant performance loss incurred by performing RA and power loading without taking into account the inter-D2D interference. In that sense, FBMC/OQAM can be classified as a disruptive technology, as it allows the management of the network to be simplified through a change in the PHY layer.

\balance

We also investigated in which scenarios the use of FBMC/OQAM is the most beneficial, identifying that it offers the greatest benefit in scenarios consisting of small, dense clusters that are located near the macro-cell edge. Identifying the scenarios in which FBMC/OQAM offers a significant benefit is of key importance. The indication that FBMC/OQAM is best suited to particular scenarios gives rise to the concept of waveform-as-a-commodity, whereby the choice of waveform is influenced by the scenario. The observations in this paper provide a platform from which to devise a network policy for determining which waveform a D2D pair should use, based on the scenario. We plan to extend this work to a multi-cell, multi-cluster case in which multiple D2D clusters may exist in the same cell, and each cluster is assigned a waveform independently. In addition, clusters in neighbouring cells would also interfere with one another in this case.

\section*{Acknowledgements} 

{\small This publication has emanated from research conducted with the financial support of Science Foundation Ireland (SFI) and is co-funded under the European Regional Development Fund under Grant Number 13/RC/2077. The work of the authors with CentraleSup\'{e}lec was partially funded through French National Research Agency (ANR) project ACCENT5 with grant agreement code: ANR-14-CE28-0026-02.}

\bibliographystyle{templates/IEEEtran}  
\bibliography{IEEEabrv,d2d_waveform.bib}

\end{document}